# An exciton-polariton laser based on biologically produced fluorescent protein

Christof P. Dietrich[1,2,*], Anja Steude[1], Laura Tropf[1], Marcel Schubert[1], Nils M. Kronenberg[1], K. Ostermann[3], Sven Höfling[1,2,*], Malte C. Gather[1,*]

[1]SUPA, School of Physics and Astronomy, University of St Andrews, North Haugh, KY16 9SS St Andrews, United Kingdom.

[2]Technische Physik, Wilhelm-Conrad-Röntgen-Research Center for Complex Material Systems, Universität Würzburg, Am Hubland, 97074 Würzburg, Germany.

[3]Insitut für Genetik, Technische Universität Dresden, 01062 Dresden, Germany.

*Correspondence to: christof.dietrich@physik.uni-wuerzburg.de, sven.hoefling@physik.uni-wuerzburg.de, mcg6@st-andrews.ac.uk .

**Abstract**:

Under adequate conditions, cavity-polaritons form a macroscopic coherent quantum state, known as polariton condensate (PC). Compared to Wannier-Mott excitons in inorganic semiconductors, the localized Frenkel excitons in organic emitter materials show weaker interaction with each other but stronger coupling to light, which recently enabled the first realization of a PC at room temperature. However, this required ultrafast optical pumping which limits the applications of organic PCs. Here, we demonstrate room-temperature PCs of cavity-polaritons in simple laminated microcavities filled with the biologically produced enhanced green fluorescent protein (eGFP). The unique molecular structure of eGFP prevents exciton annihilation even at high excitation densities, thus facilitating PCs under conventional nanosecond pumping. Condensation is clearly evidenced by a distinct threshold, an interaction-induced blueshift of the condensate, long-range coherence and the presence of a second threshold at higher excitation density which is associated with the onset of photon lasing.

**Introduction**

Strong light-matter coupling occurs when interactions between the exciton reservoir of an active medium and a surrounding confined light field - generated typically by an optical microcavity - are strong enough to result in the formation of bosonic coupled light-matter particles, known as cavity-polaritons (*1*). Under appropriate conditions, these integer spin particles relax into a common ground state with a joint particle wavefunction - a process called polariton condensation (PC) (*2,3*). PC can occur at significantly lower thresholds than regular photon lasing (*4,5*) because no population inversion is required (*6*). Owing to their limited exciton binding energy, most inorganic semiconductor materials do not support PC at room temperature. By contrast, the Frenkel type excitons found in organic materials are stable at room temperature (*7*) and show enormous coupling strengths (up to 1 eV (*8*)). So far, however, PC of 'organic polaritons' has only been reported for a handful of materials and has required optical pumping with ultrashort pulses (sub 10 ps) to achieve sufficient excitation density without excessive exciton-exciton annihilation (*9-11*). In addition, fabricating high-Q microcavities without damaging the organic emitter contained inside has necessitated elaborated device fabrication processes. Finally, the characteristic two threshold behavior expected for a polariton laser (where the second threshold indicates the onset of photon lasing) has not been observed in organic devices so far.

Here, we demonstrate polariton lasing from simple laminated high-Q microcavities filled with the biologically produced enhanced green-fluorescent protein (eGFP). The emission of these hybrid bio-inorganic devices shows a clear two-threshold behavior, corroborating the build-up of a polariton condensate at low excitation densities and the onset of photon lasing (with simultaneous transition to the weak coupling regime) at higher excitation densities. Both, polariton condensation and photon lasing are achieved with conventional nanosecond excitation, eliminating the need for ultrashort pump pulses.

There is growing interest in understanding how biomaterials affect light at the quantum level, e.g. with respect to coherence and entanglement in photosynthetic complexes (*12, 13*). Photo-synthetic complexes are also studied as the active part of polariton devices (*14*). Fluorescent proteins (FPs) form a different class of biomaterial and in the context of quantum biology offer properties, which are complementary to photosynthetic complexes. The discovery, cloning and continued engineering of FPs (*15-18*) has revolutionized biomedical imaging and FPs have become an essential tool in *in vivo* microscopy to label gene products and cellular components. Despite their widespread use, FPs were until recently used mainly in the spontaneous emission regime and in the aqueous, highly diluted environment inside cells. However, recent work has shown that they are also interesting for stimulated emission (*19-21*). In addition, due to their unique molecular geometry, FPs are efficient solid-state fluorescent emitters, unlike many synthetic dyes (*22-24*). The FP molecule consists of a nano-cylinder of 11 *β*-sheets that enclose the actual fluorophore at the center of the molecule (Fig. 1a) (*25, 26*). The *β*-sheets act as a molecular bumper that ensures a 3-4 nm separation of fluorophores, even in the highly concentrated solid-state form of the material (Fig. 1b). This separation dramatically slows down exciton diffusion and reduces concentration quenching (*27*). We hypothesized that the molecular bumper may also be beneficial to reduce exciton-exciton annihilation which otherwise affects the performance of gain media at the high excitation levels required for lasing and polariton condensation.

## Results and Discussion

### Suppression of exciton-exciton annihilation in eGFP

To investigate the performance of FPs at high exciton concentrations, we compared the fluorescence of thin films of eGFP and of 2,7-bis[9,9-di(4-methylphenyl)-fluoren-2-yl]-9,9-di(4-methylphenyl)fluorine (TDAF), one of the materials in which PC of organic polaritons was recently achieved (using fs or ps optical pumping) (*10*,*11*). When excited with low intensity nanosecond light pulses, both films initially showed a linear increase in brightness with pump fluence (Fig. 1c). However, above a critical fluence of $E_{0,\text{TDAF}} = 0.037$ mJ/cm$^2$ the brightness of the TDAF sample began to increase more slowly ($\propto E^{1/2}$). Such a square root dependence is a fingerprint of bimolecular exciton-exciton annihilation being the dominant channel of exciton decay (*28*). The eGFP film does not show significant exciton-exciton annihilation up to a twenty-fold higher pump fluence ($E_{0,\text{eGFP}} = 0.80$ mJ/cm$^2$). Using the critical fluence for each material and a rate equation model, we estimate effective annihilation rate constants of $k_{\text{XX,eGFP}} = 2.4 \times 10^9$ s$^{-1}$ and $k_{\text{XX,TDAF}} = 6.1 \times 10^{10}$ s$^{-1}$ for eGFP and for TDAF, respectively (Supplement S1), indicating that the high exciton densities required for polariton condensation are achieved more readily with eGFP. The relatively low annihilation rate constant also implies that large exciton densities may be achieved in eGFP films using nanosecond instead of picosecond or femtosecond pumping.

### eGFP filled laminated microcavities

We next fabricated laminated microcavities comprising of an eGFP film (thickness, 500 nm) sandwiched between two dielectric Bragg mirrors (Fig. 1d). The lamination process yields a one-dimensional in-plane thickness gradient of ~ 1.5 - 4.5 µm (Fig. 1e), which allows us to tune the cavity resonances across a wide spectral range. The eGFP film has a root-mean-square roughness of 8.2 nm as determined by atomic force microscopy (Fig. 1f). The absorption of a pristine protein film is dominated by two excitonic transitions located at 2.53 eV (490 nm, X1) and 2.67 eV (465 nm, X2), respectively (Fig. 1g). The emission spectrum peaks at 2.44 eV (508 nm). Both the absorption and emission spectrum are significantly broadened and show considerable overlap. Combined with the spectral and spatial overlap between cavity and photon mode, this overlap between emission and absorption is indispensable for effective energy transfer between photon and exciton and thus for polariton condensation. By contrast, large exciton linewidths have been considered as hurdle to achieving condensation (*7*) and much effort has been directed towards finding organic materials with particularly narrow linewidths so that the polariton linewidth is smaller than the Rabi splitting of the coupled system (*10*). Here, we follow a different approach: By increasing the overall cavity thickness and coupling several cavity modes to the excitonic transitions of eGFP (Fig. 1h), we boost the Q-factor of the involved cavity modes and increase the photonic character of the coupled cavity polaritons (86% photonic, Supplement S2) which then also reduces the polariton linewidth.

We found that the density of eGFP within the active region of the microcavity is crucial to realizing strong cavity-photon interaction. Reflectance measurements of microcavities containing eGFP with high and low water content (see Methods) are shown in Fig. 2. The spectra were recorded at the short wavelength edge of the stop band of the dielectric mirrors to increase the visibility of the involved effects. As a consequence, the spectra show contributions of the cavity modes (CM*n*) but also the first Bragg mode (i.e., Bragg minimum) of the dielectric mirrors (BM). For the microcavity containing eGFP with high water content, the cavity modes and the Bragg mode overlap spectrally (at certain thicknesses). By contrast, a clear splitting occurs between all observed modes and for all thicknesses for the eGFP film with low water content. This avoided crossing is a clear signature of strong coupling between eGFP excitons and cavity modes. We attribute the difference between the two samples to the higher concentration of eGFP fluorophores in the film with low water content. In 'traditional' polariton microcavities with only one exciton and one photon mode, anti-crossing is expected between upper and lower polariton branch as well as between polariton branches and uncoupled exciton and photon modes. Only the latter is present here due to the multimodality of both the photonic and electronic system. Fitting a coupled oscillator model to the experimental data for the low water content sample yields coupling strengths of $V_1 = (97\pm8)$ meV and $V_2 = (46\pm5)$ meV for the X1 and X2 transitions, respectively (Supplement S2).

### First threshold and polariton lasing

We subsequently pumped the microcavities off-resonance ($\lambda = 460$ nm, close to the X2 band) using nanosecond pulses. While the exciton lifetime of eGFP is in the few-ns-range (*22*), the lifetime of cavity photons is ≈ 200 fs (determined from the natural linewidth in a similar weakly coupled microcavity). This means that the LP2 polaritons have a sub-ps lifetime (≈ 250 fs) and therefore that the nanosecond excitation pulses effectively constitute a steady-state excitation of the system. Fourier imaging was used to record the angular-dependent microcavity emission at different excitation pulse energies (Fig. 3 and Supplement S3). At low excitation energies (< 10 nJ), emission was from the lower polariton branches LP*n* with strongly decreasing intensity towards higher emission angles, following a Boltzmann occupation along the polariton dispersion (Supplement S4), thus showing the absence of any polariton bottleneck (*10*). When increasing the excitation energy, a critical point $P_1 = 12$ nJ was reached where the emission i) shifted to higher energies, ii) increased nonlinearly in intensity (see Fig. 4a) and iii) drastically narrowed in linewidth (also see Fig. 4c) as well as in momentum spread around zero emission angle. This indicates the onset of polariton condensation by scattering of polaritons into the ground state of the LP2 dispersion. Furthermore, the polarization of the condensate is pinned to the polarization of the pump source (Supplement S5). Energetically the condensate lies well above the eGFP gain spectrum (*23*) which excludes conventional photon lasing as explanation for these observations. Finally, we observed the formation of interference fringes at condensation threshold (Supplement S6). We attribute this to the presence of a collective macroscopic phase of polaritons with spatial coherence.

When increasing the pump energy further, the polariton condensate peak shifted even more into the blue (Fig. 4b, filled circles) which is a sign of nonlinear polariton-polariton and polariton-exciton interactions in the condensate. By contrast, the lower dispersion curve does not shift considerably with pump energy (Fig. 4b, open circles) for excitation powers up to 2.5 $P_1$. This is further evidence for polariton lasing and against conventional photon lasing. (In the latter, frequency pulling of the lasing peak towards the gain maximum of eGFP would cause a redshift with increasing pump energy.) At the same time, the emission from the condensed cavity polaritons remains energetically below the expected dispersion of uncoupled cavity photons (Fig. 3b and c, dotted line); the maximum blue shift of the condensate is $(9.2\pm1.0)$ meV whereas the energy difference between the coupled and uncoupled dispersion minimum is $(15.1\pm0.5)$ meV.

**Second threshold and photon lasing**

For excitation energies above $3P_1$, we observed a shift of the emission from the coupled polariton dispersion to the uncoupled cavity photon dispersion (Fig. 3d). We interpret this as transition to the weak coupling regime due to depletion of the eGFP ground state. At a pump energy of $P_2 = 125$ nJ, the microcavity emission began to again increase super-linearly with pump energy. This second threshold is an order of magnitude higher than $P_1$ and is associated with the onset of photon lasing in the protein microcavity. Photon lasing is again accompanied by a decrease in linewidth (Fig. 4c) but there is no interaction-induced blueshift. Note that the photon lasing peak does not originate from the dispersion minimum of the uncoupled cavity mode (around 2.42 eV) associated with the mode that showed polariton condensation but is strongly redshifted to the next, energetically lower uncoupled cavity mode (at around 2.28 eV). We attribute this to mode hopping into the cavity mode closest to the gain maximum of eGFP at around 2.25-2.3 eV (*23*). For the higher-energy cavity mode, the material gain is insufficient to achieve conventional photon lasing. This observation further supports our interpretation that the first, lower threshold is associated with polariton condensation.

**Dependence of thresholds $P_1$ and $P_2$ on spectral detuning**

Changing the detuning of a particular cavity mode with respect to the excitonic transitions (by scanning along the thickness gradient of the microcavity) gives insight into the wavelength dependence of the thresholds for polariton ($P_1$) and photon ($P_2$) lasing (Fig. 4d). We observed two bands with minima at 2.40 eV and at 2.31 eV for $P_1$ and $P_2$, respectively. The $P_2$-band is a qualitative measure of the eGFP gain spectrum and agrees well with our recent investigations (*23*). The $P_1$-band clearly peaks outside of the gain region of eGFP and the lowest polariton condensation threshold is an order of magnitude smaller than the lowest $P_2$ value.

**Conclusion**

The compatibility with nanosecond optical pumping and the simplicity of our lamination approach represent major improvements over previous organic polariton lasers. They bring polariton condensation and its remarkable properties - such as low-threshold coherent light generation and superfluidity - within easier reach and represent an important step towards cw polariton lasing at room temperature. To the best of our knowledge, the two-threshold-behavior observed here has not been reported for an organic system so far due to a combination of photo-bleaching and exciton-exciton annihilation at high excitation densities (*10*, *11*). The small Rabi splitting of our system combined with the very weak bimolecular quenching make it much easier to fully deplete the ground-state and thus leave the strong-coupling regime. The observation of polariton and photon lasing from one device thus demonstrates the robustness and efficiency of eGFP as active optical material. In the future, cavities loaded with a combination of different FPs will enable controlled studies of energy transfer. By exploiting the complex interaction of FPs with water, we expect that solvent-assisted micro-shaping will yield complex intra-cavity patterns, paving the way to the study of macroscopic quantum interference effects in FP films.

**Materials and Methods:**

Sample preparation: The open reading frame of enhanced green fluorescent protein (eGFP) was PCR-amplified and cloned into vector pET23b (Novagen, Germany) using standard molecular biology methods (*30*). The corresponding protein contains an N-terminal HA-tag for immune-detection and a His(6)-tag for protein purification. *E. coli* strain BL21(DE3)pLysS (Novagen, Germany) was transformed with the construct using LB-medium with ampicillin (100 μg/ml) and chloramphenicol (34 μg/ml) for selection. Cells were grown at 30°C until OD600 of 0.5 - 0.6 and harvested 4 h after inducing protein expression by Isopropyl-beta-D-thiogalactopyranosid (IPTG; final concentration, 0.5mM). Cells were resuspended in buffer (20mM Tris pH 7.9, 500mM NaCl, 5mM Imidazol), disrupted by sonification and treatment with lysozyme, and eGFP was purified using His-Bind Resin (Novagen, Germany). After elution, the protein containing fraction was dialyzed against PBS twice utilizing a Slider-A-Lyzer Dialysis cassette (10K MWCO, 15ml) (Thermo Fisher Scientific, USA). Subsequently, eGFP was filtered and centrifuged to remove buffer salts and increase protein concentration (to up to 220 g/l). Concentrated solutions (~ 100 μl) were spin-coated (1000 rpm for 60 s) onto dielectric mirrors (distributed Bragg reflectors, surface roughness below $\lambda/10$) designed for peak reflectance ($R \geq 99.995\%$) at a wavelength of 532 nm and consisting of 14 pairs of alternating $SiO_2$ (73 nm)/$Ta_2O_5$ (59 nm) layers. The structure was capped with an identical mirror on top to form the microcavity. Subsequently, the eGFP solution dries out and leaves a solid eGFP film with a nearly constant thickness of around 500 nm. Microcavities were characterized directly after spin-coating and again after the drying process. In this way, the same sample with different water content of the eGFP film inside the cavity is investigated. The manual capping yielded a gradient in cavity thickness, spanning from 1500 to 4500 nm, which enabled scanning through different cavity modes by translating the sample. The theoretical quality factors of the photonic modes reach up to $Q$ = 50,000. Thin films of eGFP for fluorescence characterization were fabricated on glass substrates in the same manner as described above. TDAF (Lumtec) was used as a representative example of a synthetic organic polariton laser material. Thin films of TADF were deposited by thermal evaporation under high vacuum.

Cavity characterization: Reflectance measurements were performed with a custom designed microscopic setup in epi-illumination and Fourier imaging configuration. A broad band white LED was focused onto the sample surface through a high numerical aperture objective (NA = 0.42) covering an angular range of ±25°. The reflected light was collected through the same objective, dispersed by a 500 mm imaging spectrograph with a spectral resolution down to 0.03 nm and imaged onto a cooled EM-CCD array detector. For measurements of the microcavity emission, the structure was excited by a pulsed, wavelength-tunable OPO system tuned to 460 nm (pulse length, 7 ns) using the same optical setup as for the reflectance measurements. The pump laser is slightly elliptic and is focused down to an 8 μm spot (average FWHM) on the sample. The same system was also used for the fluorescence measurements of eGFP and TDAF on glass substrates.

Matrix simulations: In order to model the experimentally observed dispersions of cavity photons and polaritons and to determine the coupling strengths between eGFP excitons and cavity photons, we calculated the eigenstates of a coupled oscillator matrix that accounts for several cavity modes strongly interacting with two main excitonic transitions of the eGFP fluorophore. The matrix reads as follows

$$\begin{pmatrix} E_{X2} & 0 & V_2 & V_2 & \cdots & V_2 \\ 0 & E_{X1} & V_1 & V_1 & \cdots & V_1 \\ V_2 & V_1 & E_{Ph,1} & 0 & \cdots & 0 \\ V_2 & V_1 & 0 & E_{Ph,2} & \cdots & 0 \\ \vdots & \vdots & \vdots & \vdots & \ddots & \vdots \\ V_2 & V_1 & 0 & 0 & 0 & E_{Ph,n} \end{pmatrix}$$

with the main exciton transition $E_{X1}$ at 2.53 eV, its less pronounced sideband transition $E_{X2}$ at 2.67 eV and the uncoupled photon mode energies $E_{Ph,n}$ as well as the coupling strengths $V_1$ and $V_2$ to the transitions $E_{X1}$ and $E_{X2}$, respectively. Due to the simultaneous coupling to different cavity modes CM$n$ and the Bragg mode BM, we treat the light field as a non-pure state (superposition of different modes). For most spectra, a good quantitative agreement between experiment and theory was achieved by taking ten photon modes close to the exciton resonances into account. The uncoupled cavity dispersions were determined by transfer matrix calculations (Supplement S2) using a background refractive index of 1.51 for the spin-coated eGFP thin film.

**Acknowledgments:** We thank A. Clemens (TU Dresden, Germany) for technical support with protein preparation and C. Murawski (U St Andrews, UK) for support with TDAF deposition.

**Funding:** We acknowledge support from the ERC Starting Grant ABLASE (640012), the Scottish Funding Council (via SUPA), the European Union Marie Curie Career Integration Grant (PCIG12-GA-2012-334407), studentship funding through the EPSRC CM-DTC (EP/L015110/1) and the EPSRC Hybrid Polaritonics program grant (EP/M025330/1). S.H. gratefully acknowledges support by the Royal Society and the Wolfson Foundation and M.S. gratefully acknowledges support from a MSCA IF (659213).

**Author contributions:** C.P.D. fabricated the protein microcavities, and characterized them together with N.M.K. and L.T.. K.O. was responsible for providing fluorescent proteins, A.S. and M.S. prepared the protein solutions. M.C.G. and S.H. conceived the project and guided the experiments. M.C.G. designed the structures. All authors contributed to the analysis of the data. C.P.D. and M.C.G. co-wrote the manuscript with input from all authors.

**Competing interests.** The authors declare that they have no competing interests. All data needed to evaluate the conclusions in the paper are present in the paper and/or the Supplementary Materials. Additional data available from authors upon request.

**Supplementary Materials:**

S1. Analysis of pump fluence dependent fluorescence measurements

S2. Transfer matrix and coupled oscillator matrix calculations

S3. Excitation-dependent zero-momentum emission of the eGFP microcavities

S4. Thermalization of the polariton emission

S5. Polarization of the condensate

S6. Spatial coherence of the polariton condensate

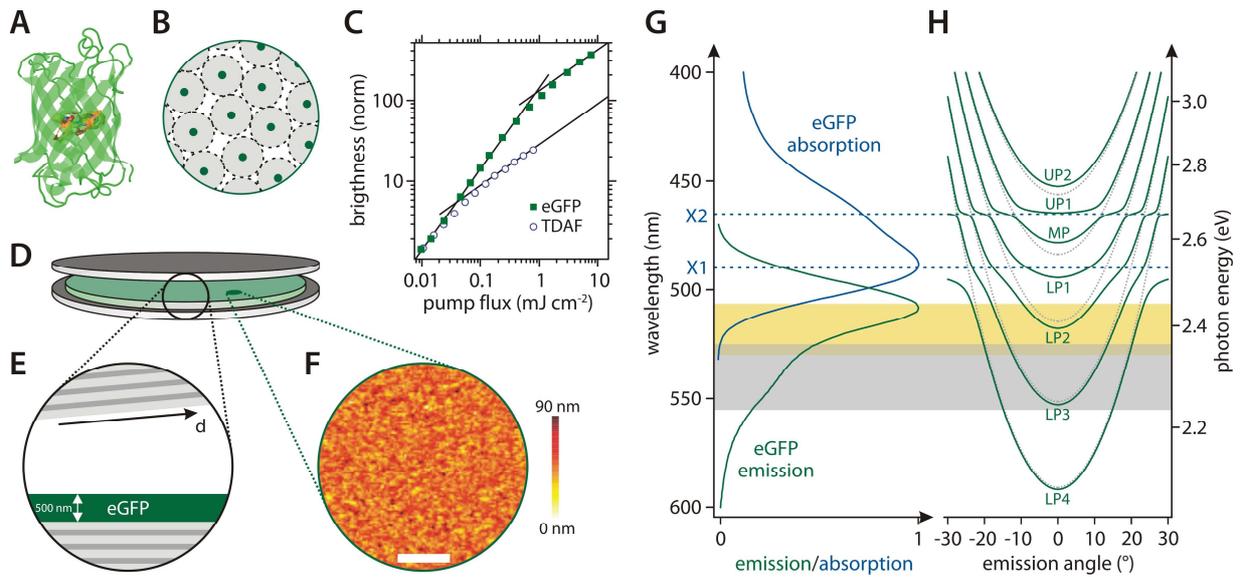

**Fig. 1**. Laminated eGFP-filled high-Q microcavity used for polariton lasing. (A) Molecular structure of eGFP with the fluorophore in the center surrounded by 11 β-sheets. (B) Schematic of how the β-sheets prevent concentration quenching. (C) Normalized brightness versus pump fluence of thin films of solid-state eGFP and of the synthetic organic polariton material TDAF. Bi-molecular exciton-exciton annihilation reduces spontaneous emission at high pump fluence (linear intensity increase changes to sub-linear increase with slope 1/2). eGFP tolerates twenty-fold higher pump fluence before the annihilation-induced sub-linear behavior sets in. (D) Schematic illustration and (E) cross-section of the microcavity containing a 500 nm-thick film of solid state eGFP. The top mirror is slightly wedged with respect to the bottom mirror to allow adjustment of the cavity resonances by tuning the total cavity thickness $d$. (F) Atomic force microscopy topography map of the spin-coated GFP layer. Scale bar, 5 μm. (G) Absorption and emission spectrum of a solid state eGFP film. The blue dashed lines indicate the two pronounced exciton resonances. The grey area marks the region over which gain is observed in eGFP (from Ref. (*23*)), the yellow region marks the range of polariton lasing (see Fig. 3 and 4). (H) Dispersion relation for uncoupled photonic modes (grey dotted lines) and strongly coupled polariton modes (green solid lines), calculated for typical cavity dimensions using a transfer matrix algorithm.

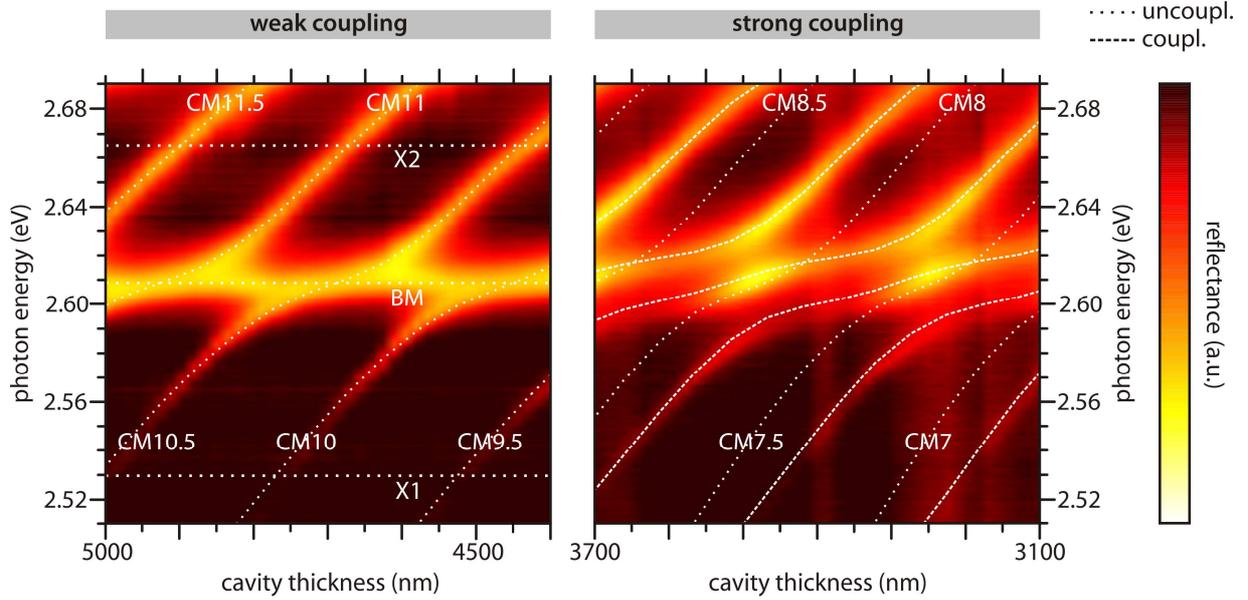

**Fig. 2.** Experimentally obtained thickness-dependent reflectivity spectra (transverse-electric polarized) for eGFP filled microcavities. Dotted lines represent uncoupled photon (CM*n*, BM) and exciton modes (X1, X2). Dashed lines represent eigenvalues of the coupled oscillator matrix obtained from coupled harmonic oscillator calculations; associated coupling strength $V_1$ = 97 meV and $V_2$ = 46 meV. In the weak coupling regime (left, high water content), modes cross whereas strong exciton-photon interaction (right, low water content) shows anti-crossing of the involved modes.

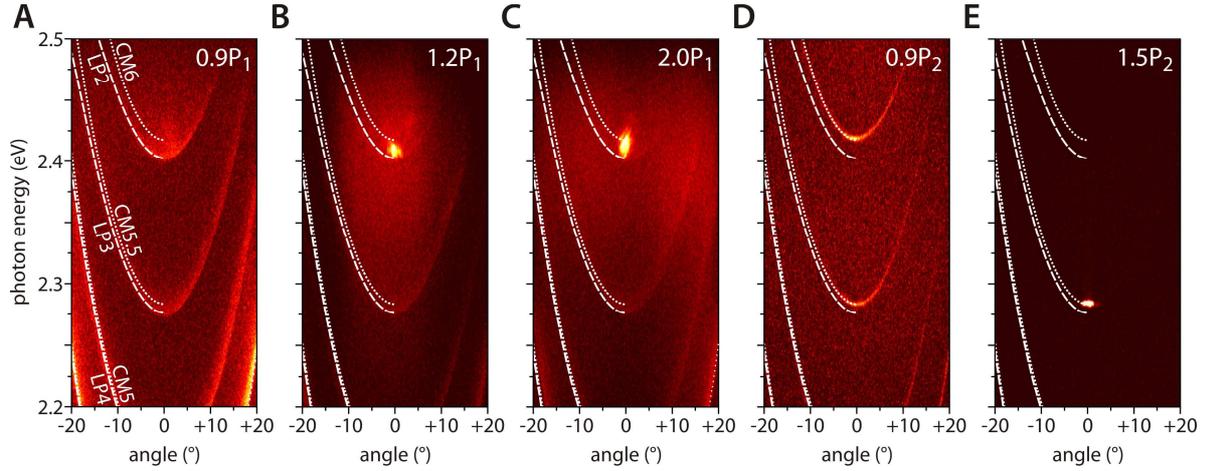

**Fig. 3.** Excitation energy dependent, angle resolved luminescence spectra of low water content eGFP filled microcavity. (A) Excitation below polariton lasing threshold ($P = 0.9P_1$). Expected position of lower polariton branches LP2, LP3 and LP4 (white dashed lines) and cavity modes CM5, CM5.5 and CM6 (white dotted lines) are shown. Note that emission is present along the polariton branches and to a lesser extent blueshifted from the dispersion minimum of the LP2 branch. (B) At excitation energies above $P_1$, a distinct peak (attributed to the polariton condensate) emerges at an emission angle of 0° and is blueshifted relative to the minimum of the LP2 dispersion. (C) At $P = 2P_1$, the blueshift of the condensate-peak has increased further. (D) Around $P = 9P_1 = 0.9P_2$ the emission into the LP modes disappears and the polariton peak collapses. Instead, emission from the uncoupled photon modes CM$n$ occurs. (E) At $P = 1.5P_2$, a sharp emission peak appears at 0° from the CM5.5 mode.

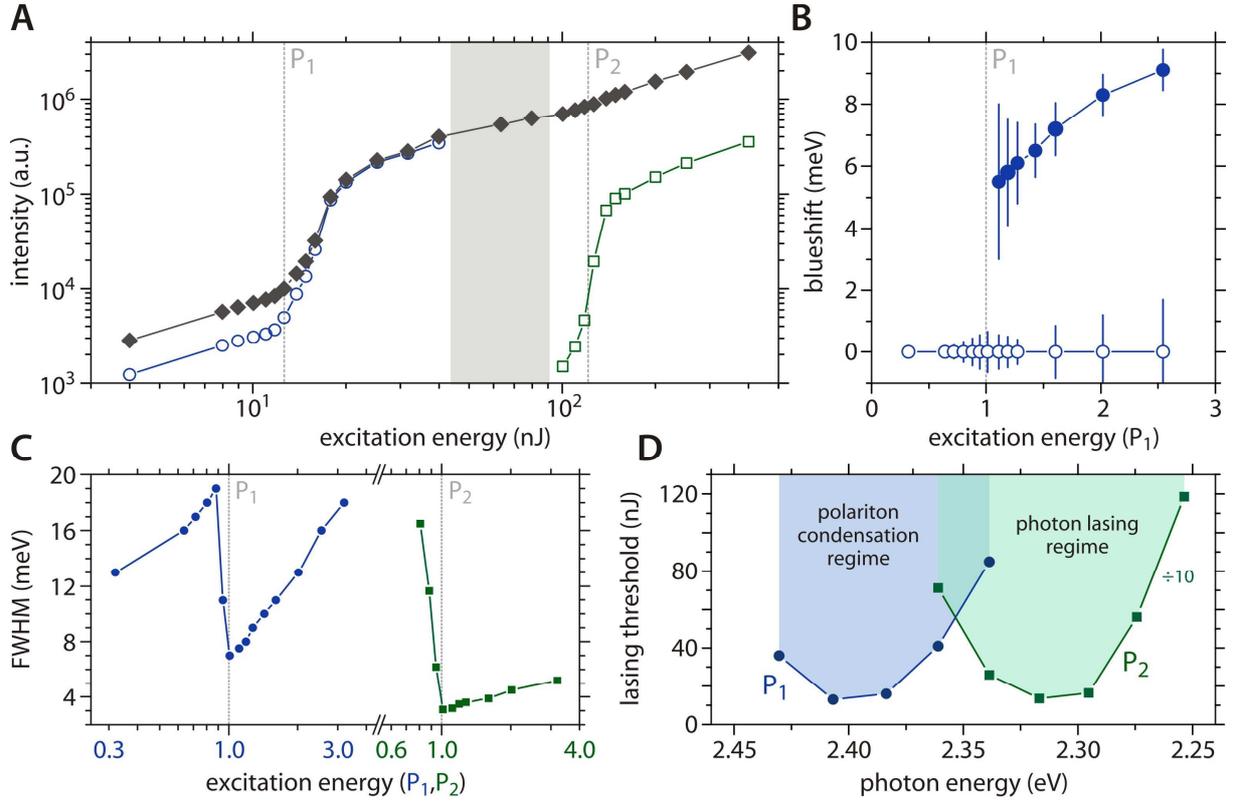

**Fig. 4.** Analysis of condensate and photon laser properties. (A) Integrated microcavity emission intensity (filled diamonds) as well as individual contributions of LP2 (open circles) and CM5.5 mode (open squares) versus excitation energy. The two thresholds are attributed to the onset of polariton lasing ($P_1$) and photon lasing ($P_2$). (B) Spectral shift of the polariton condensate (closed symbols) and LP2 polariton branch (open symbols) as function of the excitation energy (in units of $P_1$). (C) Spectral width of the LP2 polariton emission (blue symbols) and the photon emission around CM5.5 (green symbols) as a function of excitation energy (in units of $P_1$ and $P_2$, respectively). (D) Threshold energies of polariton lasing ($P_1$) and photon lasing ($P_2$) for different tuning of the cavity. Cavity tuning is achieved by scanning along the thickness-gradient of the microcavity. $P_2$ is rescaled for clarity. Note the two distinct minima in threshold for the polariton condensation and photon lasing regime, respectively (2.40 eV and 2.31 eV).

# An exciton-polariton laser based on biologically produced fluorescent protein

## Supplementary Materials

Christof P. Dietrich[1,2,*], Anja Steude[1], Laura Tropf[1], Marcel Schubert[1], Nils M. Kronenberg[1], K. Ostermann[3], Sven Höfling[1,2,*], Malte C. Gather[1,*]

[1]SUPA, School of Physics and Astronomy, University of St Andrews, North Haugh, KY16 9SS St Andrews, United Kingdom.

[2]Technische Physik, Wilhelm-Conrad-Röntgen-Research Center for Complex Material Systems, Universität Würzburg, Am Hubland, 97074 Würzburg, Germany.

[3]Insitut für Genetik, Technische Universität Dresden, 01062 Dresden, Germany.

*Correspondence to: christof.dietrich@physik.uni-wuerzburg.de, sven.hoefling@physik.uni-wuerzburg.de, mcg6@st-andrews.ac.uk .

### S1. Analysis of pump fluence dependent fluorescence measurements

Let $n$ be the density of fluorophores in a thin film of fluorescent material, $n_1(t)$ the density of excited state fluorophores and $n_0(t) = n - n_1(t)$ the density of non-excited fluorophores. In general, the temporal evolution of $n_1$ can be described by

$$\frac{dn_1(t)}{dt} = \frac{I_p(t)\sigma}{h\nu_p} n_0(t) - kn_1(t) - \gamma n_1(t)^2, \quad (1)$$

where $I_p(t)$ and $\nu_p$ are the intensity and frequency of the pump light, $h$ is Planck's constant, $\sigma$ is the absorption cross section of the fluorophores, $k = 1/\tau$ is the decay rate or inverse excited state lifetime of the fluorophore, and $\gamma$ is the bi-molecular quenching constant.

We numerically solved Eq. 1 for a pump pulse with a Gaussian temporal profile, a FWHM pulse duration of $t_p = 7$ ns and pump fluences between $10^{-6}$ and $10^{-1}$ J/cm². The fluorescence intensity at each point in time is proportional to $n_1(t)$. Integrating $n_1(t)$ over the entire length of the pulse thus provides a measure of the total fluorescence generated. Fig. 1 shows the dependence of $\int n_1(t)dt$ on pump fluence $E$. One can clearly distinguish the region of linear increase at lower fluence (here, emission is the dominant channel by which excitons decay) and a region with square root type increase at higher fluence (here, the dominant channel of exciton decay is through exciton-exciton annihilation). The pump fluence $E_0$ at which the slope changes is the point where the rates for radiative decay and exciton-exciton annihilation are equal, i.e.

$$kn_1(t) = \gamma n_1(t)^2. \quad (2)$$

To estimate the exciton-exciton annihilation rate constant from $E_0$, we solved Eq. 1 analytically which can be readily done if one assumes constant pump light intensity, i.e. $I_p(t) = I_{p,0}$. The steady state limit of this analytic solution is given by

$$\lim_{t\to\infty} n_1(t) = \frac{1}{2\gamma}\left(\sqrt{\frac{4I_{p,0}\sigma}{h\nu_p}\gamma n + \left(\frac{I_{p,0}\sigma}{h\nu_p} + k\right)^2} - \left(\frac{I_{p,0}\sigma}{h\nu_p} + k\right)\right). \quad (3)$$

Using the same argument about the fluorescence being proportional to $n_1(t)$ as above, $\lim_{t\to\infty} n_1(t)$ also provides a measure of the generated fluorescence. Furthermore, as shown in Fig. S1, $\lim_{t\to\infty} n_1(t)$ derived from the analytical solution and $\int n_1(t)dt$ from the numerical solution show nearly identical dependence on pump fluence, if $I_{p,0} = E/t_p$ is taken as steady state pump light intensity. Taking $E = E_0$ and inserting Eq. 3 into Eq. 2, we can therefore obtain the effective annihilation rate constant $k_{XX}$ of the material

$$k_{XX} = \gamma n = \frac{3}{4}\left(\frac{E_0\sigma}{2t_p h\nu_p}\right)^{-1}\left(\frac{E_0\sigma}{2t_p h\nu_p} + k\right)^2. \quad (4)$$

This expression only depends on the pump fluence $E_0$, at which exciton-exciton annihilation becomes dominant, on the wavelength and duration of the pump pulses and on the absorption cross section and excited state lifetime of the fluorophore. The latter two are known for eGFP and for the TDAF molecule studied here; they can also be measured relatively easily for most other materials. Using for eGFP, $\tau_{eGFP}$ = 3.3 ns and $\sigma(\lambda=470$ nm$) = 2.1\times10^{-16}$ cm$^2$ (Ref. (22)), we obtain $k_{XX,eGFP} = 2.4\times10^9$ s$^{-1}$. For the TDAF used as a reference, we have $\tau_{TDAF}$ = 0.89 ns and $\sigma(\lambda=355$ nm$) = 1.4\times10^{-15}$ cm$^2$ (Ref. (29)) giving $k_{XX,TDAF} = 6.1\times10^{10}$ s$^{-1}$.

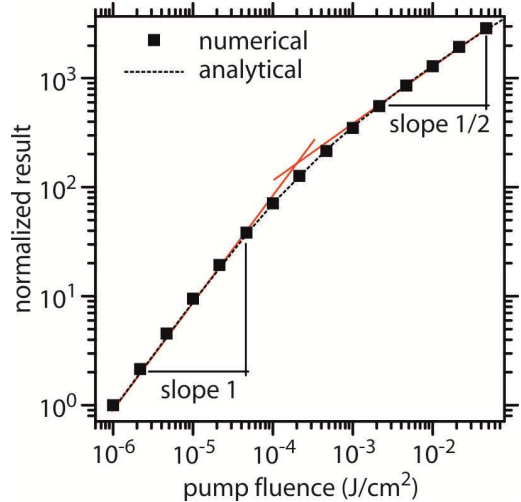

***Fig. S1.*** *Modelling of bi-molecular exciton-exciton annihilation. Dependence of numerical solution of Eq. 1 (squares) and of analytical solution for steady state case (dashed black line) on pump fluence. Red lines are guides to the eye.*

## S2. Transfer matrix and coupled oscillator matrix calculations

In order to model the uncoupled photonic system formed by the highly reflective dielectric mirrors, transfer-matrix calculations of a microcavity filled with a passive, non-absorbing material were performed, i.e. the material had no excitonic contribution. The sample structure consisted of a 1mm-thick $SiO_2$ substrate with 14 pairs of alternating $SiO_2$ (73 nm) and $Ta_2O_5$ (59 nm) layers on top forming the bottom dielectric mirror, followed by a 500-nm-thick eGFP layer, an air gap with variable thickness and an identical dielectric mirror on top. The central wavelength of the mirrors was $\lambda = 532$ nm. The overall cavity thickness is the sum of the thicknesses of the eGFP layer and the gap above. The refractive index of eGFP was assumed to be $n_{eGFP} = 1.51$.

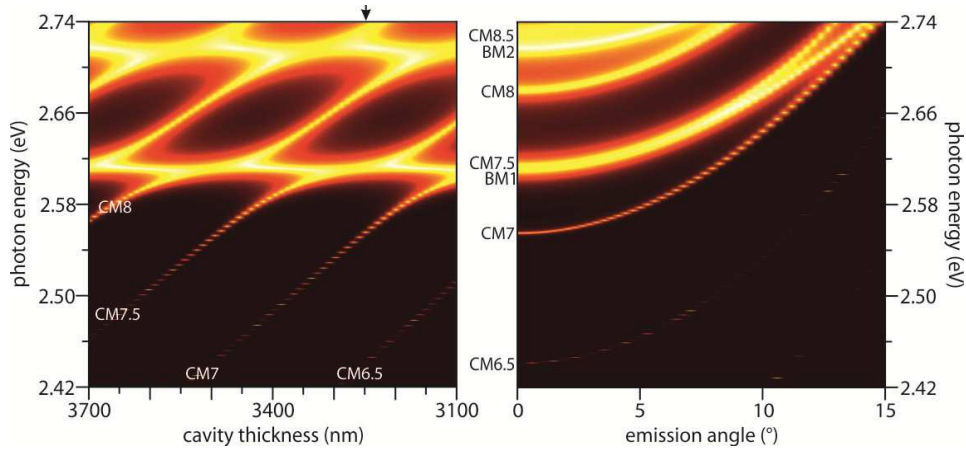

**Fig. S2**. *Reflectance of a passive microcavity. Thickness- (left) and angular-dependent (right) calculated reflectance of a microcavity consisting of identical top and bottom dielectric mirrors (14 pairs $SiO_2/Ta_2O_5$, 532 nm central wavelength), a passive non-absorbing eGFP layer (500 nm thickness) and an air gap of varying thickness. The black arrow indicates the cavity thickness assumed for the angular-resolved reflectance map, d = 3240 nm.*

The reflectance of this structure was calculated as a function of the overall cavity thickness and the emission angle. The results are shown in Fig. S2 (the thickness-dependent reflectance is shown for normal incidence). Cavity modes show mode quality factors up to $Q = 50,000$. By changing the cavity thickness, the spectral position of the confined cavity modes shifts across the stop band of the dielectric mirrors since the conditions for constructive interference change. By contrast, the Bragg modes of the dielectric mirrors are not subject to these shifts as their position is given by the structure of the mirror itself. Since the mirror distance is a multiple $n$ of the central wavelength of the dielectric mirrors, Fig. S2 shows several cavity modes with different mode numbers CM$n$. Note that cavity modes can have half-integer mode numbers since constructive interference between the dielectric mirrors arises for every $\lambda/2$.

The angular-dependent reflectance of the passive eGFP microcavity shows a blueshift of all photonic resonances with increasing angle because of their photonic dispersions. Note that the Bragg modes (BM$n$) show a less pronounced blueshift, which is due to their different photonic confinement.

In order to extract the excitonic and photonic fractions of exciton-polaritons, we determined the Hopfield coefficients $n$ by using the following equation:

$$\begin{pmatrix} E_{X2} & 0 & V_2 & V_2 & \cdots & V_2 \\ 0 & E_{X1} & V_1 & V_1 & \cdots & V_1 \\ V_2 & V_1 & E_{Ph,1} & 0 & \cdots & 0 \\ V_2 & V_1 & 0 & E_{Ph,2} & \cdots & 0 \\ \vdots & \vdots & \vdots & \vdots & \ddots & \vdots \\ V_2 & V_1 & 0 & 0 & 0 & E_{Ph,n} \end{pmatrix} \begin{pmatrix} \alpha \\ \beta \\ \gamma_1 \\ \gamma_2 \\ \vdots \\ \gamma_n \end{pmatrix} = E_{LPn} \begin{pmatrix} \alpha \\ \beta \\ \gamma_1 \\ \gamma_2 \\ \vdots \\ \gamma_n \end{pmatrix} \quad (5)$$

with $E_{X1}$, $E_{X2}$ being the excitonic energies, $V_1$, $V_2$ the coupling constants, $E_{ph,n}$ the uncoupled photon energies and $E_{LPn}$ the eigenvalues of the matrix. The squares of the entries of the eigenvectors ($\alpha$, $\beta$, $\gamma_1$, …, $\gamma_n$) are the exciton/photon content for each polariton eigenstate. For our system, we limited the number of cavity modes to $n = 10$.

Figure S3 shows the calculated excitonic fractions $\alpha^2$ and $\beta^2$ as a function of cavity thickness (left) and emission angle (right). With increasing cavity thickness, the cavity modes shift to smaller energies, thus affecting the coupling mechanism. As a consequence, the polariton states also shift to smaller energies. By changing the cavity thickness, the respective polariton eigenvalues in Fig. S3 were tuned across the entire spectral range of interest (from 550 nm to 450 nm). As an example, one of the photonic fractions ($\gamma_7^2$) is also shown in Fig. S3 (black line). The excitonic fractions $\alpha^2$ and $\beta^2$ exhibit two maxima that are related to the excitonic transitions X1 (at 3750 nm) and X2 (at 3600 nm). When tuning the cavity to the optimal thickness for condensation (around 3500 nm in this case), the excitonic fractions are smaller indicating that the polariton condensate has a substantial photonic character. The photon fraction of the condensed polaritons is ~86% at zero emission angle. Increasing the emission angle (Fig. S3, right) increases the excitonic fraction of the polariton mode, which is due to the polariton dispersion-induced blueshift of the lower polariton branch and the subsequent crossing of the excitonic transitions X1 and X2.

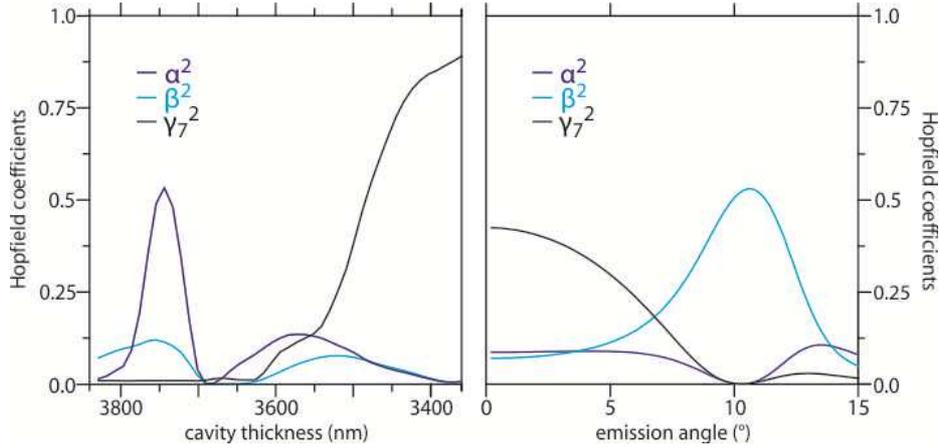

***Fig. S3.*** *Hopfield coefficients. Thickness- (left) and angular-dependent (right) calculated Hopfield coefficients $\alpha^2$, $\beta^2$ and $\gamma_7^2$ of the microcavity described above. The Hopfield coefficients were calculated by using Eq. 5 with coupling coefficients $V_1 = 97$ meV and $V_2 = 46$ meV. The angular-dependent Hopfield coefficients represent excitonic and photonic fractions of the polariton state LP2.*

*S3. Excitation-dependent zero-momentum emission of the eGFP microcavities*

Figure S4 shows the non-normalized spectra of the microcavity emission at zero emission angle for different excitation powers. Like the angular-resolved luminescence maps shown in Fig. 3 of the main manuscript, the different spectra show the condensation of the polariton mode LP2 (16 nJ, blue line), the interaction-induced blueshift of the condensate (25 nJ, grey line), the disappearance of the condensate at excitation powers >50 nJ and the subsequent onset of the weak coupling regime (110 nJ, red line) as well as photon lasing (198 nJ, light grey line).

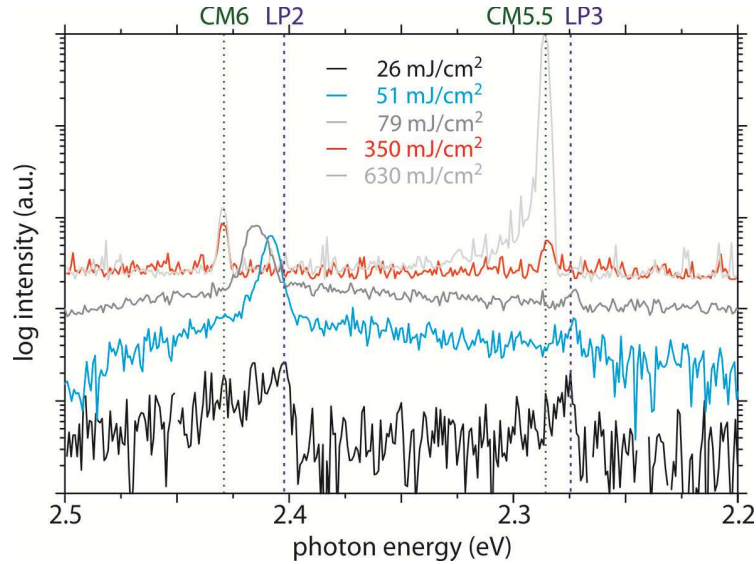

***Fig. S4**. Zero-momentum emission of eGFP filled microcavity under nanosecond optical excitation. Excitation powers are given in the upper left corner. The spectral positions of polaritonic (LPn) and photonic (CMn) transitions are indicated by vertical dashed and dotted lines, respectively.*

Note that the derived pump pulse energy threshold value for the onset of polariton lasing ($P_1 = 12$ nJ) relates to a peak excitation power density of 3.4 MW/cm$^2$ and an excitation flux of 24 mJ/cm$^2$ (taking into account an excitation spot of 8 μm and pulse lengths of 7 ns).

## S4. Thermalization of the polariton emission

Fig. S5 shows the distribution of the polariton occupancy along the dispersion LP2 for different pump fluences (c.f. Fig. 3 in main text for definition of LP2). The occupancy is corrected for the energy-dependence of the density of states at different momentum values (also accounting for the change of polariton character along the dispersion) and is normalized. The data was then fitted using a Boltzmann distribution

$$I \propto \exp\left(-\frac{E_{LP2} - E_{LP2}(k=0)}{k_b T}\right) \ .$$

This yielded effective polariton temperatures of 757 K, 373 K and 315 K for excitation densities 0.1 $P_1$, 0.9 $P_1$ and 1.0 $P_1$, respectively. For excitation near the condensation threshold, the determined polariton temperature is close to the room temperature value, meaning that polaritons are fully thermalized. However, we emphasize that even close to condensation threshold the polariton occupancy rather follows a Maxwell-Boltzmann-like behavior than a Bose-Einstein distribution (indicated by the absence of Bose narrowing close to the dispersion minimum). From this, we conclude that our system is in a non-equilibrium state *(31,32)*. Thermal equilibrium for a polariton condensate was experimentally observed in long-lived microcavities *(33)*.

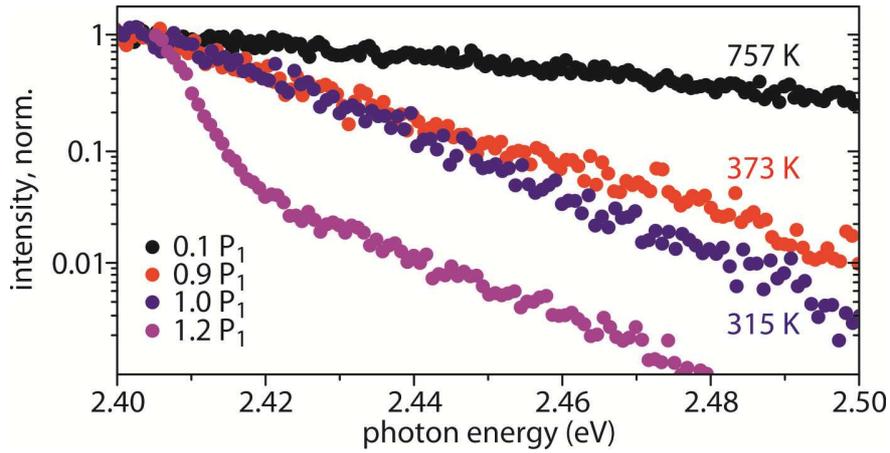

***Fig. S5.*** *Energy-dependence of occupation of the LP2 polariton branch for different excitation densities. Boltzmann-fits to the data reveal temperatures of 757 K, 373 K and 315 K for excitation densities of 0.1 $P_1$, 0.9 $P_1$ and 1.0 $P_1$, respectively. Above threshold (1.2 $P_1$), the intensity is drastically enhanced towards to the ground state (at 2.4 eV) while maintaining the same high-energy tail as for P = 1.0 $P_1$.*

## S5. Polarization of the condensate

Bose-Einstein condensation is accompanied by symmetry breaking which manifests itself in a condensate polarization that is pinned to the polarization of the excitation laser. In order to verify the process of polarization pinning, we examined the polarization-dependent response of the microcavity filled with eGFP under fixed polarization of the excitation beam (parallel to the entrance slit of the spectrometer, 0°→180°). While the microcavity emission shows no polarization-dependence at all in the linear regime ($P < P_1$; Fig. S6, left), a clear pinning is observed for the polariton condensate ($P_1 < P < P_2$; Fig. S6, right). Since the dipoles of the eGFP fluorophores have a random orientation, only fluorophores with a significant dipole moment parallel to the linearly polarized excitation laser are excited during pumping. As pointed out in Ref. (*10*), depolarization due to Förster energy transfer is much slower than the onset of condensation leading to the preservation of the excitation polarization in the condensate. Note that the polarization dependence in Fig. S6 is different from Fig. 3 in the main manuscript where we excited the microcavity with unpolarized light and detected the emission only for TE-polarization.

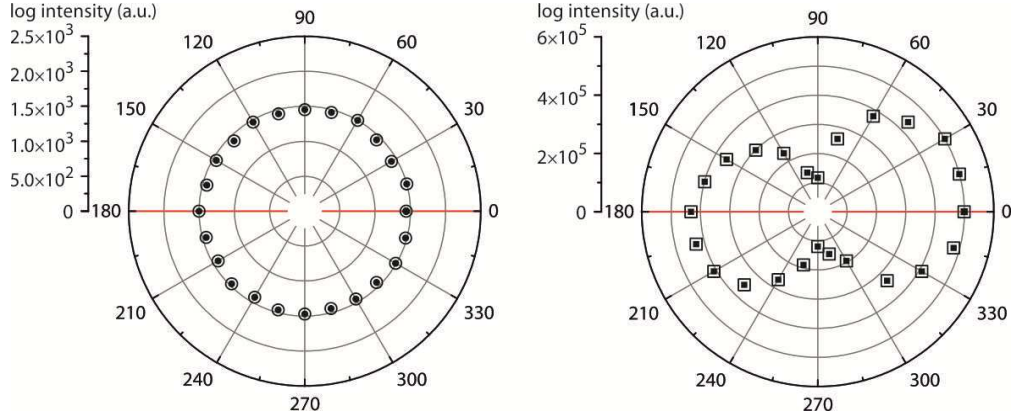

***Fig. S6.*** *Polarization pinning of the condensate. Polarization-dependent integrated intensity of the microcavity emission below condensation threshold (left) and above (right). The red line indicates the linear polarization plane of the excitation laser.*

*S6. Spatial coherence of the polariton condensate*

We measured the spatial coherence of the polariton condensate by using a Michelson interferometer in mirror-retro-reflector configuration: the real-space luminescence image from the microcavity is split by a 50:50 beam splitter into two arms. One arm is directed towards a 45° retro-reflector flipping the image and reflecting it back to the beam splitter. The other arm is directed towards a plane mirror that is mounted on a piezo-controlled linear translation stage. In this way, the image of the sample and its reversed image interfere with each other at the output of the beam splitter where a high-resolution imaging CCD camera records the incident interference patterns. The position of the plane mirror is tuned to maximize fringe contrast. Examples of the resulting interference patterns are shown in Fig. S7, for excitation powers below (A) and well above condensation threshold $P_1$ but still below photon lasing threshold (B). The images are single-shot measurements (excited by a single 7 ns pulse at 460 nm) and the emission is passed through edge pass filters to only show the 500-600 nm range. The laser spot is indicated by white dashed ellipses in Fig. S7 (half the peak power of the Gaussian excitation beam). Whereas the pattern does not show any significant structure below threshold, clear fringes are evident above threshold providing evidence for the presence of coherence across the spatial extend of the condensate. Note that in Fig. S7B, the lateral extension of the interference fringes is wider than the excitation spot. Since the system is pumped well above threshold and the dashed line indicates half the peak power of the excitation beam, we assign this to coherence contributions from sample areas that are still pumped above threshold but are outside the marked area. [Given the 0.55 numerical aperture of the used imaging system, we believe that the increase in size is unlikely to reflect a limit of the point-spread function of the imaging system *(34)*.] Furthermore, the interference fringes occasionally show fork-like dislocations (see white circle in Fig. S7).

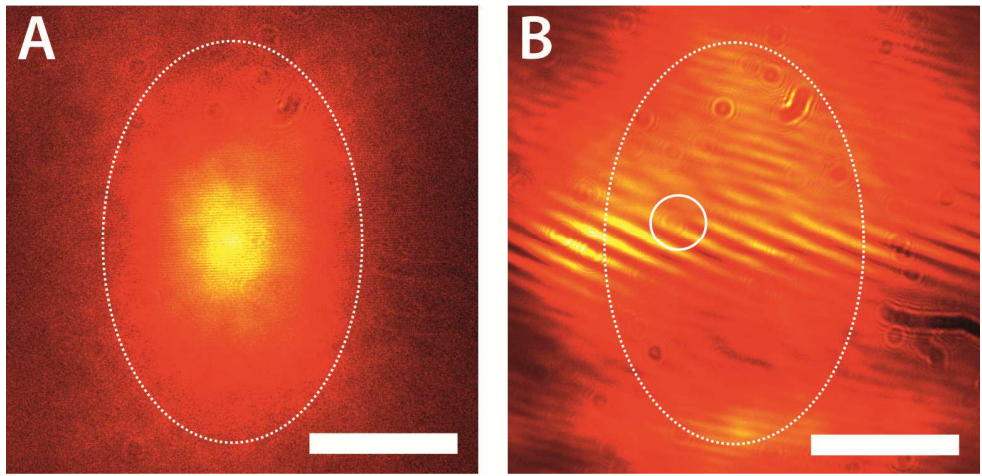

***Fig. S7.*** *Spatial coherence of eGFP polariton condensates. Single-shot, real-space Michelson interferograms of an eGFP microcavity excited below (a) and above (b) condensation threshold $P_1$. The image is spectrally filtered by using 500 nm long-pass and 600 nm short-pass filters. The white circle in the right image indicates a fork-like dislocation of the interference fringes. Scale bars, 5 μm.*